\documentclass[12pt]{article}
\usepackage{latexsym}
\usepackage{epsfig}
\usepackage{graphicx}
\usepackage[english]{babel}
\usepackage{epstopdf}
\usepackage{epsfig}

\begin{document}
\title{Large t diffractive $J/\psi$ photoproduction with proton dissociation
in ultraperipheral pA collisions at LHC.}
\author {
L. ~Frankfurt\\
\it School of Physics and Astronomy,\\
\it  Tel Aviv University,
Tel Aviv, 69978 , Israel\\
M. ~Strikman\\
\it Department of Physics,
Pennsylvania State University,\\
\it University Park, PA  16802, USA\\
 M.~Zhalov\\
\it St. Petersburg Nuclear Physics Institute, Gatchina, 188300
Russia}
\date{}
\maketitle

\begin{abstract}
We evaluate
the large momentum transfer $J/\psi$ photoproduction 
with rapidity gaps in ultraperipheral proton-ion collisions at the LHC which provides an effective method of probing dynamics of 
large t elastic hard QCD Pomeron interactions.
 It is shown that the experimental studies of this process
would allow  to investigate the energy dependence of cross 
section of elastic scattering of a small $c\bar c$ dipole off the gluon  over
 a wide range of invariant energies  
$10^3 <  s_{c\bar c - gluon} < 10^6 \, GeV^2$. The accessible 
energy range exceeds  the one reached in $\gamma p$ at 
HERA by a factor of 10 and allows the kinematic cuts which
improve greatly 
sensitivity to the Pomeron dynamics as compared to
the HERA measurements.
  The cross section is expected to change 
by a factor $\ge 20$ throughout this interval and our estimates 
predict quite reasonable counting rates for this
process with the several of the LHC detectors.

\end{abstract}
\newpage

\section{Introduction} 
\label{intro}
\bigskip

For many years investigations of the small $x$ dynamics were focused on studies of
the $\gamma^{*}(\gamma ) p$ interactions ar HERA. In particular,
over the last decade significant efforts were  aimed on the
theoretical and experimental investigations of the high energy  diffractive
photo/electro production of vector mesons off proton target (for the recent review and references see 
\cite{nikolaev})
including the hard processes
in which a vector meson is produced with a  large rapidity gap,
$\Delta y$,  between the vector meson $V$ and the produced hadronic system $X$:
\begin{equation}
\gamma^* + p\to V + \, gap \, 
+ X.
\end{equation}
An important feature of these processes
which occur 
 due to the elastic scattering of small color singlet quark-antiquark
($q\overline q$) dipole configurations in the photon wave function 
(referred to as dipoles in the remainder of this
article) off a parton within the  proton target is the nontrivial
interplay between evolution in $\ln(x_0/x)$ and $\ln(Q^2/Q_0^2)$. 
 Study of high energy vector meson  photoproduction in kinematics  where momentum transferred through 
the gluon ladder exchange between dipole and a parton 
is large and the produced vector meson is separated by large rapidity gap from 
the products of target dissociation should allow
to investigate $\ln (x_0/x)$ evolution at fixed momentum transfer $t$ \cite{AFS,FR95,Bartels}.  
The HERA measured the relevant cross sections of $\rho$ and $J/\psi$ photoproduction \cite{Derrick:1995pb,Adloff:2002em,Chekanov:2002rm,Aktas:2003zi,Aktas:2006qs} in a rather restricted  
$\gamma p$ center-of-mass range $20 \leq W_{\gamma p}\leq 200$ GeV and very limited range of $\Delta y$, making it very difficult
to study the energy dependence of the amplitude predicted by the BFKL dynamics.
Hence, it would clearly be beneficial to perform such measurements at 
higher $W_{\gamma p}$ and over a much larger range of  $\Delta y$.

Here we extend our  feasibility studies of  
\cite{FSZ06}, \cite{rholrg}  for probing these 
processes in the proton-nucleus ultraperipheral collisions
(UPCs)\cite{yelpage}
 at the LHC. The CMS and ATLAS detectors are 
well suited for such investigations since they cover large rapidity intervals. 
The ALICE detector maybe capable of studying this process in
a certain rapidity range as well.

We consider $J/\psi$ photoproduction by photon from the ultrarelativistic ion
on the proton target in ultraperipheral $pA$  collisions,
\begin{equation}
A+p\to A+\gamma + p \to A+J/\psi + X \, \,.
\label{eqvm}
\end{equation}
in the kinematics with large $t$ and the rapidity gap $\Delta y $ 
between $J/\psi$ and the produced hadronic system $X$ which is
sufficiently large to 
suppress the fragmentation contribution. 

The paper is organized as follows. In section \ref{formula} we
discuss the QCD motivated formulas and parametrizations which 
could be used to describe the energy and $t$ dependence of the considered process. 
 Section \ref{hera} is devoted to analysis of the available HERA data on $J/\psi$ production 
using  parametrization 
of the data which is based on the hard mechanism of the reaction.
In section \ref{rates} we estimate the rates 
of $\gamma + p \to J/\psi + \, gap +\, X$  reaction in $pA $ collisions 
at LHC.

\section{Energy and $t$ dependence in the large $t$ and rapidity gap quarkonium photoproduction}
\label{formula}

The main variables in such processes are the mass of the system produced in the proton 
dissociation, $M_X$, the square of momentum transfer $-t\equiv Q^2= 
-(p_{\gamma}-p_{V})^2$, and the square of the $q\overline q$-parton elastic 
scattering energy
\begin{equation}
s_{(q \overline q)j}=xW_{\gamma p}^2 = x s_{\gamma p} \, \,.
\end{equation} 
Here $j$ denotes gluon or quark in the target and
\begin{equation}
x=\frac {-t} {(-t+M_X^2-m_N^2)} \, \,, 
\label{xmin}
\end{equation}
is the fraction of the proton momentum carried by the 
target parton for a given $M_X$ and $t$.  
At large $t$ and $W_{\gamma p}$, the gap between the rapidity of
the produced vector meson and the final-state parton at the leading edge of the rapidity 
range of the hadronic system $X$ is
\begin{eqnarray} 
\Delta y = \ln \frac {xW_{\gamma p}^2} {\sqrt{(-t)(M_{J/\psi}^2-t)}} \, \, .
\label{gapdef}
\end{eqnarray}
It is rather difficult to measure $M_X$ or $x$ 
directly.  However, they can be adequately determined by studying
the leading hadrons close to the rapidity gap; full reconstruction
is not required.

 In kinematical region of large $-t \gg 1/r_{J/\psi}^2 $ and
$W_{\gamma p}^2\gg M_X^2$  the quarkonium photoproduction 
cross section with target proton dissociation can be described as an 
incoherent sum of terms which are proportional to the product of the 
large $t$ cross section of quarkonium photoproduction off the parton 
and the density of the parton $j$ in the target 
\cite{AFS,FR95}:  
\begin{eqnarray}
\frac {d\sigma_{\gamma p\to J/\psi X}} {dt dx}=
\frac {d\sigma_{\gamma q \to J/\psi q}} {dt} \biggl [{81\over 16} g_{p}(x,t) 
+\sum_i [q_{p}^{i}(x,t)+{\overline q}_{p}^{i}(x,t)] \biggr ] \, \, .
\label{DGLAPBFKL}
\end{eqnarray}
Here $g_{p}(x,t)$, $q_{p}^{i}(x,t)$ and $\overline q_{p}^{i}(x,t)$ are the 
gluon, quark and antiquark distributions in the proton. 
At large $t$ the $\gamma q \rightarrow J/\psi \, q$ amplitude, 
$f_q (s_{(c \overline c)q} ,t)$, is dominated by
transition of the photon into the small $c\overline c$ dipole
configuration which scatters elastically off the target parton and 
transforms into the vector meson. 
The dipole - gluon elastic scattering cross section 
is enhanced by a factor
81/16 relative to the scattering off quark.  
Large $t$ ensures important simplifications. The 
parton ladder mediating elastic scattering of dipole is attached 
via two gluons to one target parton while the attachment 
to two and more partons in the target is strongly suppressed.  
Since $t$ is the same on all rungs of the ladder, 
the amplitude $f_{q}(s_{(c \overline c)j},t)$ probes evolution in 
$\ln(1/x)$ at fixed $t$
\cite{FS, MT}. 
Because the momentum transfer is shared between two gluons, the
characteristic virtuality of $t$-channel gluons on the ladder is 
$\approx -t/4$ while the hard scale in the target parton density is 
$\approx -t$.

To the lowest order in $\ln(1/x)$, the amplitude, 
$f_{q}(s_{(c \overline c)j},t)$, is independent of $W_{\gamma p}$ for fixed  $t$. 
 Higher order terms in $\ln (1/x)$ accounted for
in the leading and next-to-leading log approximations 
will result in increase of $f_q$ with energy as a power of 
$\exp(\Delta y)$ 
\begin{eqnarray}
f_q(s_{(c \overline c)j},t)\propto R(t)
\biggl (\frac {s_{(c \overline c)j}} {|t|}\biggr )^{\delta_0 +\delta^\prime t}, \, \, 
\label{dipamp}
\end{eqnarray}
for $|t| \gg M_V^2$. 

The value  $\delta_0$ changes significantly between LO and NLO BFKL,  
$\delta_0 \sim 0.6$ at LO and $\delta_0 \sim 0.1$ at NLO.  
The resummed BFKL gives a value of $\delta_0 \sim 0.2 - 0.25$ \cite{Salam:2005yp} over a wide 
range of $\alpha_{s}(Q^2)$.
In the perturbative QCD the slope parameter 
$\delta^\prime$ should be negligible ($\sim c/t $) in the $-t\to \infty$ limit.
Since kinematics of HERA measurements is restricted by the interval 
 $2 \mbox{GeV}^2\le -t \le 30 \mbox{GeV}^2$
we keep this parameter to  reveal sensitivity of the process to its value.
Experimental information about $\delta(t)=\delta_0 +\delta^\prime t$ 
in hard processes can be extracted, in principle, from the study of the exclusive 
electroproduction of light vector mesons  
or photo and electroproduction
 of heavy quarkonia.  The current $J/\psi$ data  leads to 
$\delta_0\sim 0.2 $ for a rather wide range of $Q^2$. 
The data on the $t$ dependence of $\delta(t)$ 
in electroproduction processes 
at $\left<Q^2\right> = 8.9 \mbox{GeV}^2$ are 
consistent with  $t$ independent $\delta$,  though the error bars  are significantly  
larger in this case, for the recent results see \cite{Levy:2005ap,Aktas:2005xu}. 
Therefore a natural guess for the value of $\delta (t)$ in the hard regime is 
\begin{equation}
\delta(t)\approx 0.1-0.25
\label{guess}
\end{equation}
for the $t$ interval $2 \mbox{GeV}^2\le -t \le 30 \mbox{GeV}^2$.
In view of the theoretical uncertainties described above  we will 
treat $\delta_0$ and $\delta^\prime$ as the free parameters
but generally assume that $\delta(t)$ weakly depends on $t$.

The $t$-dependence of $d\sigma_{\gamma q\to J/\psi q}/dt $  originates both 
from the square of the energy dependent part  of  elastic parton - parton scattering 
amplitude as given by Eq. \ref{dipamp} and from the t-dependence  of the attachments 
of the ladder to vector meson and parton of the target, $R^{2}(t)$. 
 The large $-t\gg M_V^2$ behavior of the non-spin flip photoproduction cross section 
is  $R^{2}(t)\propto 1/t^4$. The spin- flip contribution leads to   
 $R^{2}(t)\propto \vert 1/t\vert ^3$ with strong sensitivity 
 to the form of a vector meson-photon coupling.
 The HERA data on $J/\psi$ photoproduction \cite{Aktas:2003zi}
indicate that the spin flip contribution remains a small correction 
in the whole studied range of t. 
Hence we will neglect it 
in the following. 

The important feature of the large rapidity gap  $J/\psi$ photoproduction 
in a wide range of $t$, 
starting from $-t \sim $ few GeV$^2$, is that the essential virtualities 
in the attachments of the ladder to  the $\gamma \to J/\psi$ vertex and to the 
parton of the target are different. 
In the  $\gamma \to J/\psi$ coupling all virtualities are at least of the order $m_{J/\psi}^2$.
 This  was first observed in \cite{Forshaw} where the process of $J/\psi$ production 
was considered in the leading log BFKL approximation. 
Actually,
this feature of the process 
is more general and 
follows from 
the gauge invariance. 
The t-dependence in the ladder -target parton block should be  similar to that
in deep inelastic scattering off the corresponding parton of the target -
two gluons attached to this parton in the amplitude  have 
large relative momenta and effectively act as one point-like probe of virtuality $t$. 
Experimentally scaling in DIS sets in at $Q^2 \sim 1- 2 GeV^2$. 
Hence, to describe the $t$-dependence in the $J/\psi$ photoproduction
 we need in principle two scale parameters - 
one close to the square of the mass of $J/\psi$ and another, $t_0$, of the same 
 order as the scale of onset of scaling in DIS.

\section{Rapidity gap  $J/\psi$ photoproduction at HERA}
\label{hera}

The experimental cuts in the rapidity gap $J/\psi$ photoproduction HERA data
(\cite{Chekanov:2002rm}, \cite{Aktas:2003zi}) resulted in the
relatively small rapidity interval available for gluon emission in the color 
singlet ladder, $\ln (xs_{\gamma p}/|t|) \leq 5$. 
Really, in most of the probed settings  the invariant energy of 
the dipole-parton system was rather low,
$W_{\bar q q-j} \le 20 \,GeV$.  Since only single 
gluon emission is allowed  for such energies in the ladder kinematics, 
it is hardly justified
to 
apply a BFKL-type approximation. 
Hence, we need some reasonable QCD motivated 
parametrization of the cross section
 based on Eq.\ref{DGLAPBFKL}.
 It should 
 account for the discussed above 
two scales characterizing onset of hard processes in the large rapidity gap
$J/\psi$ photoproduction and should
 be applicable 
 in
 a wide region of $t$, not 
only to asymptotically 
large $t$. We find that the simplest choice is
:
\begin{equation}
{\frac {d\sigma _{\gamma +p\to J/\psi +Gap +X}} {dt}}=
\frac {CI(x_{\rm min},t)} {( t_{0}-t)
(M_{J/\psi}^2-t)^{3}}\cdot \biggl [ {\frac {W_{\gamma p}^2} {\sqrt{(t_{0} -t)
(M_{J/\psi}^2 -t})}}\biggr ]^{2\delta (t)}. 
\label{gjpsi}
\end{equation}

The factor $I(x_{\rm min}, t)$, is obtained by integrating the parton
densities over x,
\begin{equation}
I(x_{\rm min},t) =
\int \limits_{x_{\rm min}}^{1} dx \, x^{2\delta (t)}
 \left[\frac {81} {16} g_{p}(x,t)+
\sum _{i}[q_{p}^i (x,t)+{\overline q}_{p}^i (x,t)] \right]. \, \, 
\label{intx}
\end{equation}
To calculate $I(x_{\rm min}, t)$ we 
used 
the CTEQ6M PDFs \cite{cteq6}.
The value of $x_{min}$ was 
calculated using Eq.(\ref{xmin}) separately for each of experimental 
sets of the ZEUS and H1 data
using reported cuts on $M_X$. 
The values of $\delta_0$, $\delta^{\prime}$ in $\delta (t)=\delta_{0}+{\delta}^{\prime}t$
and constant $C$ were adjusted to provide a 
reasonable  description of the $J/\psi$ photoproduction at HERA.

\begin{figure}
\begin{center}
\epsfig{file=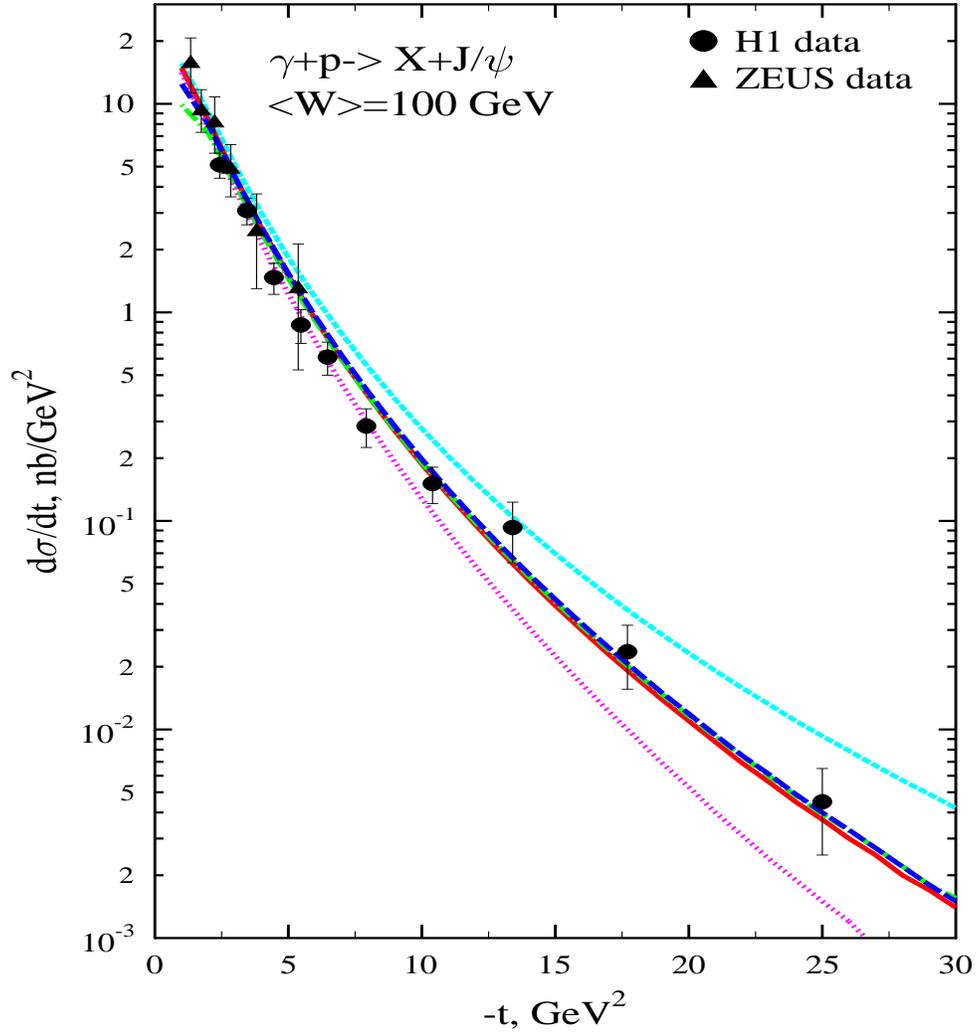, height=7in,width=6in}
 \caption{Momentum transfer distribution for $J/\psi$ photoproduction
in H1 and ZEUS. Solid-long-dashed curve corresponds to the choice of $\delta_0
=0,\,0.1,\,0.2$ with $\delta^\prime =0.01$. Short-dashed curve - 
$\delta^\prime =0.0$ and dotted curve $\delta^\prime =0.02$. Here we determine
the parameter $\delta^\prime$ and the normalization constant $C$ for each value of $\delta_0$.}
 \label{h1psit}
\end{center}
 \end{figure}

\begin{figure}
\begin{center}
\epsfig{file=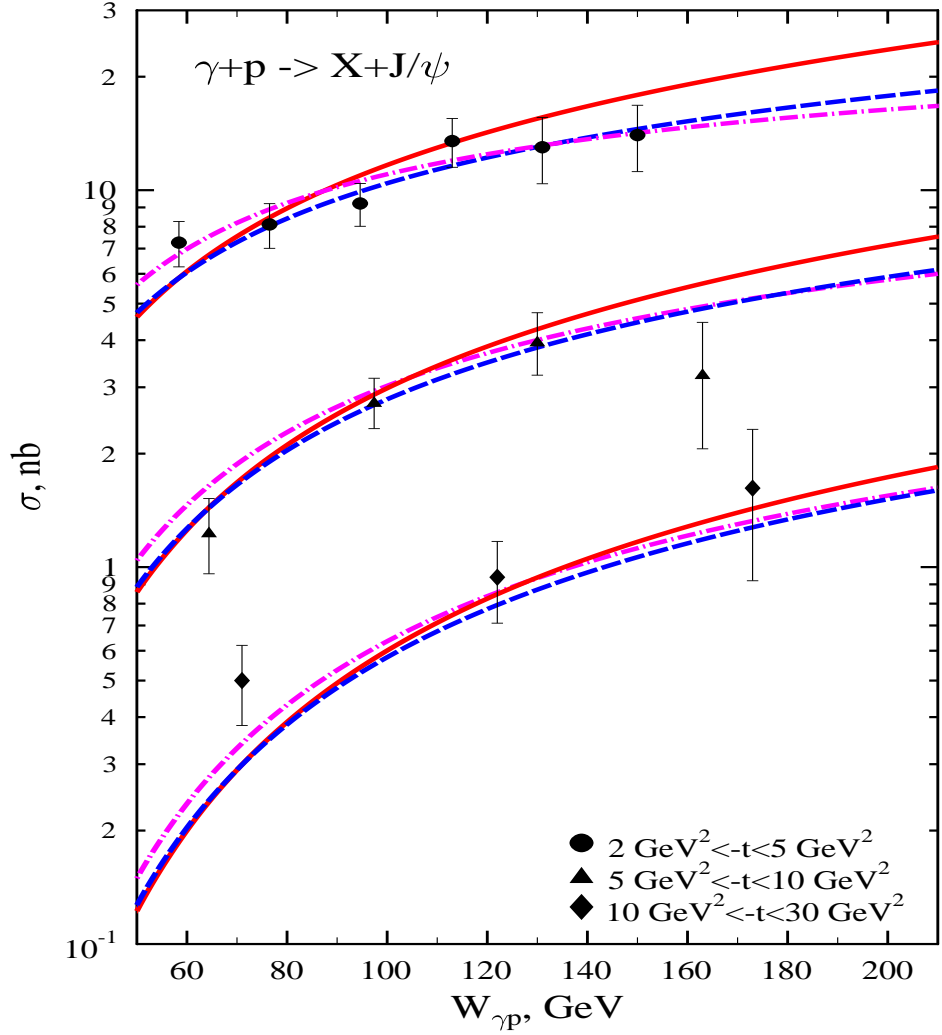, height=7in,width=6in}
 \caption{Energy dependence (choice of $\delta_0$) for $J/\psi$ photoproduction
in experiment of H1 at HERA. No free parameters used in calculations. 
Solid line corresponds to 
$\delta_0=0.2$, dashed line - $\delta_0=0.1$ and dot-dashed line
 $\delta_0=0.$ }
 \label{h1psiw}
\end{center}
\end{figure}

\begin{figure}
\begin{center}
\epsfig{file=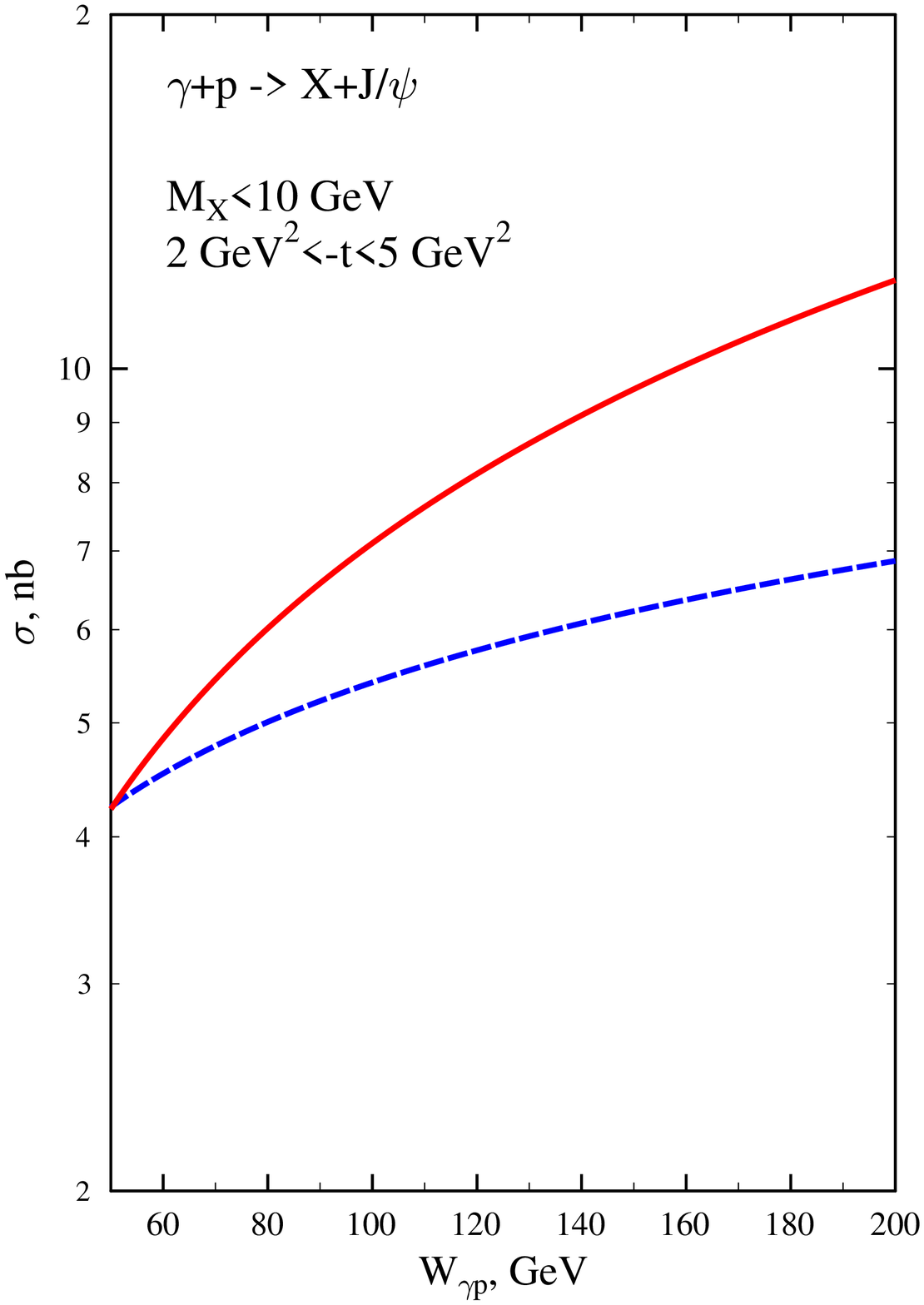, height=6in,width=5in}
 \caption{Energy dependence for $J/\psi$ photoproduction
 at HERA in kinematics of fixed upper limit of the diffractively
 produced mass $M_{max}=10 \,GeV$. Solid line corresponds to $\delta_0=0.2$
and dashed $\delta_0=0.1$. 
These calculations demonstrate that
 measurements of cross section in such kinematics could allow one 
to determine reliably the parameter  
$\delta_0 $. 
}
 \label{h1psiwfix}
\end{center}
\end{figure}

There are exists two sets of the large $t$ and rapidity gap $J/\psi$ 
data . Collaboration ZEUS \cite{Chekanov:2002rm} measured the t-dependence
of cross section in the range $1 \,GeV^2 < -t<8\,GeV^2$ at the
average energy $<W_{\gamma p}>\approx 100 \,GeV$ using the following
cuts: 
$x_{min}=\frac {-t} {W_0^2}$ with $W_0=25\, GeV$
if $x_{min} > 0.01$ otherwise
$x_{min}=0.01$. In the experiment of H1 collaboration \cite{Aktas:2003zi} 
the t-dependence has been
measured in the wider interval $2\,GeV^2<-t<30\,GeV^2$ averaging the
cross section over the energy in the range 
 $50\,GeV \le W_{\gamma p} \le  150\, GeV$ and assigning the averaged
cross section to the value $W_{\gamma p}\approx 100\,GeV$. 
The
restriction on the mass of the produced system $M_X$,
 $M^2_{X} \leq 0.05 W_{\gamma p}^2+t$, applied in H1 experiment leads to
 the energy  dependent value of $x_{min}$. This means that the cross
 section has to be integrated over x in the interval $x_{min}<x<1.$ 
before averaging over energy at each value of $t$.

We find that these
data are consistent (fig.\ref{h1psit}) with $t$ dependence predicted by QCD dominance of one 
gluon exchange in the charmonium wave function ($d\sigma /dt \propto t^{-4}$ ) and 
the shape of the curve is not sensitive to the 
parameter $\delta_0$ which characterizes explicit energy dependence 
of the amplitude of dipole scattering off a constituent of the target. 
The $t$ dependence of cross section is described 
reasonably well with the values:  $C=0.9$ if $\delta_0=0.$, 
$C=0.4$ if the parameter $\delta_0=0.1$ and $C=0.2$ for
$\delta_0=0.2$ at not too large values of $t$. Since the parameter $\delta_0$
is appeared to be comparatively small it is evident that 
there is definite sensitivity to the slope parameter 
$\delta^{\prime}$ at large $t$. From comparison with the data 
the value $\delta^{\prime}=0.01$ GeV$^2$
seems to be reasonable. 

The H1 collaboration also  presented the data
for the energy dependence of cross section for three intervals of $t$. 
In this kinematics dependence of cross section on $W_{\gamma p}$ is determined 
by the factor $W^{4\delta(t)}_{\gamma p}$ in 
the cross section of dipole-parton interaction and by integral over $x$ of 
weighed by factor $x^{2\delta(t)}$ the parton distributions  within the proton.
With an increase of the energy the value of $x_{min}$ is decreasing and due
to the strong growth of the gluon distributions in the proton the integral
over $x$ rapidly grows.  From comparison of the results of calculations with 
the data we found that in the kinematics covered by HERA, i. e. with the energy
dependent cut on the mass $M_X$ of the produced system, increase of cross section with energy
is due to the growth of $I(x_{\rm min},t)$ related to the increase of the
gluon distributions rather than due to the factor $W^{4\delta(t)}$. 
So, we conclude that the kinematical cut imposed in the H1 experiment 
results in a weak sensitivity of the data to actual value of  $\delta(t)$, see Fig.\ref{h1psiw}.

One can achieve a better sensitivity to $\delta_0$ if the cross section 
were  studied as a function of energy at fixed $M_X$ (fixed $x$) or for  
$M_X\le \, const$.  In the latter 
case the integral of parton distribution over $x$  is constant and the  energy dependence
follows from the amplitude of dipole-parton interaction. For the illustration we present
in Fig.\ref{h1psiwfix} results of the calculations for such $M_X$ cut   in the 
kinematics of H1. 

Thus  our analysis shows that  H1 and ZEUS data  on large t large rapidity gap processes  
are consistent with the 
$t$-dependence expected in pQCD which is a combination of $t$-dependence due to the structure 
of the $q\bar q$ vector meson wave function and the t-dependence of the 
 amplitude of dipole-parton scattering. 
At the same time data are insensitive to the energy dependence of the amplitude 
of dipole-parton scattering which can not  be reduced to parton distributions in a hadron.  
 To probe pQCD predictions for this amplitude one need to perform measurements in different
kinematics and at significantly larger collision energies which can be achieved in ultraperipheral 
collisions at LHC.

\section{Rapidity gap $J/\psi$ photoproduction in ultraperipheral $pA$ collisions 
at the LHC.} 
\label{rates} \bigskip

It is unlikely  that further HERA studies will cover a sufficiently wide 
range of $W_{\gamma p}$ and $\Delta y$ to study the 
energy dependence of the large-$t$ elastic dipole-parton 
scattering amplitude. On the other hand, at the LHC, CMS and ATLAS will have 
sufficient rapidity coverage to study the process in Eq.~(\ref{eqvm}) 
in ultraperipheral $pA$ collisions.
  To estimate the large $t$ rapidity gap 
$J/\psi$ 
photoproduction cross section in ultraperipheral $pA$ 
collisions at the LHC
we use the parametrization of the $\gamma p\to X J/\psi$ cross section in
Eqs.~(\ref{gjpsi}).
${ }$ We do not address the $pA$ contribution from $\gamma A\to J/\psi 
X $ since it is 
much smaller
and can easily be separated experimentally.
The large $t$ nucleon-dissociation cross section is then
\begin{equation}
{\frac {d\sigma_{pA\to J/\psi X A}} {dt dy}} = \frac{dN_{\gamma}^{Z}(y)}{dk}
{\frac {d\sigma_{\gamma N\to J/\psi +rap\, gap + X}(y)}{dt}},\, \, 
\end{equation}
where $dN_{\gamma}^{Z}(y)/dk$ is the photon flux generated by the ion
with energy $k = (M_{J/\psi}/2)\exp(y)$. We consider intermediate and large 
momentum transfer in UPCs at the LHC, analogous to those
studied at HERA.

 To check feasibility 
of such measurements and their 
 sensitivity to the value of $\delta$ we performed calculations of 
the cross sections of rapidity gap events with $J/\psi $ in ultraperipheral
proton-ion collisions at LHC for the interval 
of the $J/\psi$ rapidities $-4\le y\le 4$ which
corresponds to the range of energies $W_{\gamma p}$ from about 40 GeV
up to 1 TeV. Results are also shown for two assumptions of $\delta_0$: 0.2 and 0.1.
We do not consider $W_{\gamma p} < 20$ GeV where
our HERA-based parametrization, Eqs.~(\ref{gjpsi}) and (\ref{intx}), is
unreliable.

 \begin{figure}
\begin{center}
\epsfig{file=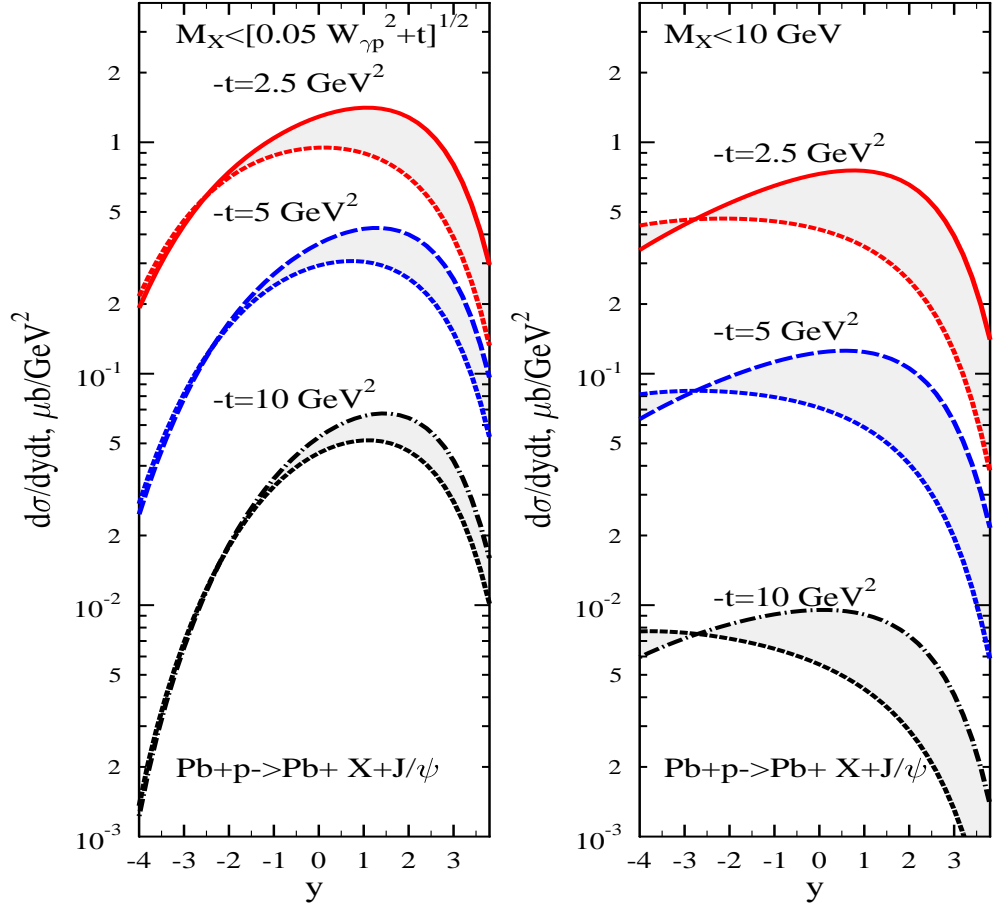, height=6.25in,width=5.25in}
 \caption{Rapidity distributions for the $J/\psi$ photoproduction in
 the ultraperipheral proton-Pb collisions at LHC in kinematics
with the gap between $J/\psi$ and diffractively produced system with the
mass in the range $m_N < M_X \le \sqrt{0.05 W_{\gamma p}^2 +t}$ (left)
and in the range $m_N < M_X \le 10\,GeV$ (right) 
at different values of the momentum transfer.
The solid, long dashed and dot-dashed lines present cross sections calculated with $\delta_0=0.2$
and the short dash lines with $\delta_0=0.1$.}
 \label{pbpfixt}
\end{center}
\end{figure}

 \begin{figure}
\begin{center}
\epsfig{file=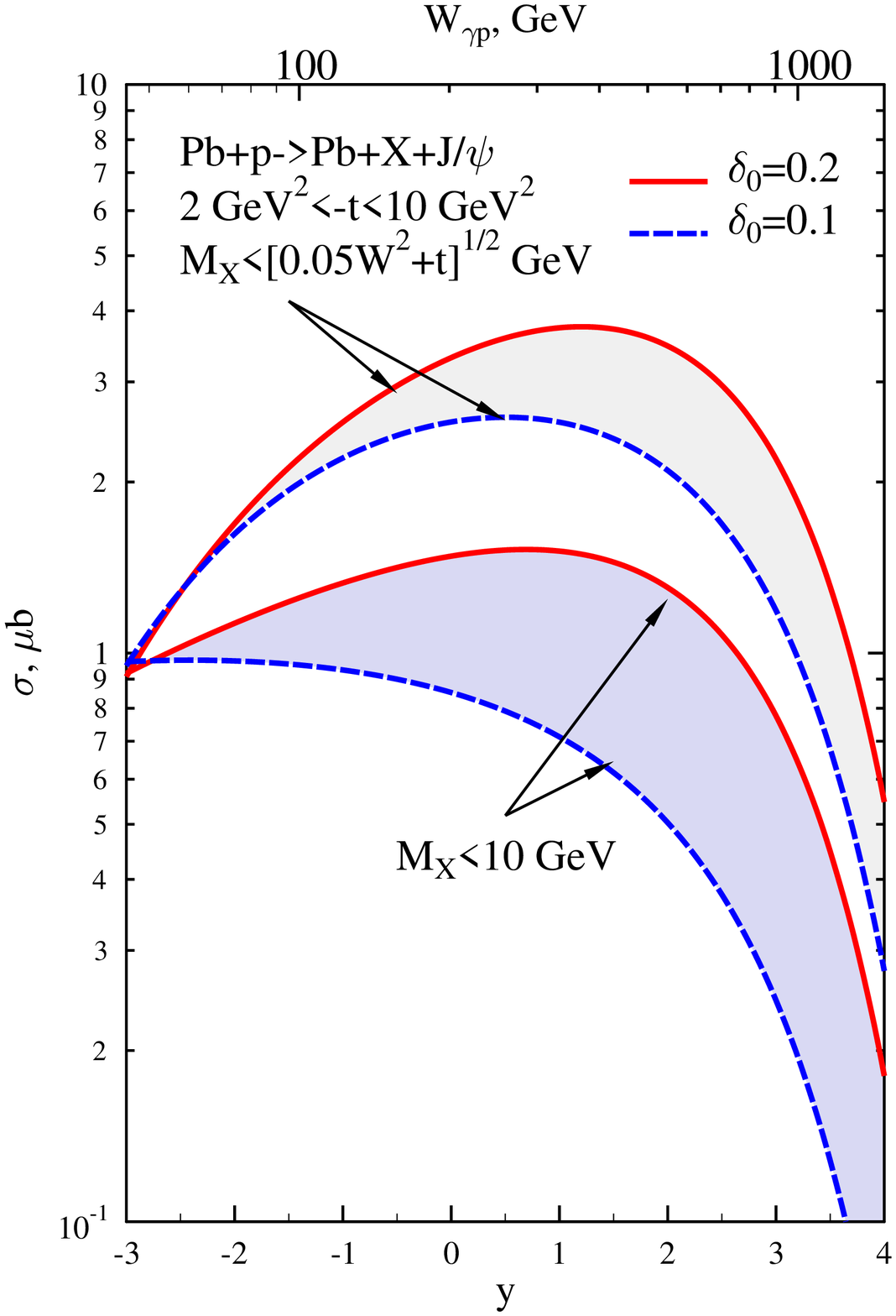, height=6.5in,width=5.5in}
 \caption{Rapidity distributions for the $J/\psi$ photoproduction in
 the ultraperipheral proton-Pb collisions at LHC in kinematics
with the gap between $J/\psi$ and diffractively produced system with the
mass in the ranges $m_N <M_X < \sqrt{0.05 W_{\gamma p}^2 +t}$ and $m_N <M_X < 10\,GeV$.
The cross sections are integrated  over $t$
in the range $2\, GeV^2 \le -t\le 10\,GeV^2$.}
 \label{pbppsiw}
\end{center}
\end{figure}


We  study the cross section when $M_X \propto W_{\gamma p}$, specifically
${M}_{X}^2 \leq 0.05\, W_{\gamma p}^2+t$ at fixed $t$, Fig.\ref{pbpfixt}(left). 
This cut corresponds to fixing $\Delta y$ and changing $x_{\rm min}$. 
Such studies could test the parton distribution 
functions and the reaction mechanism by extracting $I(x_{\rm min},t)$ from the
data in different $x_{\rm min}$ and $t$ bins. 

The cross section can also be studied at fixed $t$ as a function of the vector meson
rapidity with the restriction $M_X\leq const $ to determine the energy 
dependence of the dipole-parton amplitude and thus $\delta(t)$. 
In this case, $x_{\rm min}$ does not depend on $W_{\gamma p}$
and the dipole-parton elastic scattering amplitude
varies with $W_{\gamma p}$ due to the increase of the rapidity gap with $y$.
Most events in the kinematics with $M_X \le 10$ GeV and $2\, GeV^2 \le-t\le 10 \,GeV^2$ correspond to 
$x\ge 0.01$ where the scattering of dipole off 
the gluons of the target give the dominant contribution.
Because of the specifics of the LHC detectors it will be probably very difficult
to reach the $x$ range where quark scattering is larger than gluon scattering,
$x\ge 0.4$.  Thus we can primarily infer the energy dependence of the elastic 
$(c \overline c)$-gluon amplitude at different $t$.
  Overall, the energy range, $s_{\rm max}/s_{\rm min} \ge 
10^3$, is large enough for precision 
measurements of the energy dependence of the amplitude.
If $\delta(t) \approx 0.2$, the elastic cross section should increase by a 
factor of $\sim 30$ in the energy range.

The two choices of $M_X$ cut exhibit the same behavior at large
forward rapidity due to the steep decrease of the photon flux.      

The cross section integrated over $t$ in the range $2\, GeV^2\le -t\le 10\, GeV^2$
are presented in Figs.\ref{pbppsiw}. The total cross sections
integrated over the interval of rapidities $-4\le y\le 4$ for two considered
cuts on the mass of produced system $M_X$ are given in Table \ref{tcs}. 
The rates,
which can be obtained multiplying the cross sections by luminosities of 6 and 0.4
$\mu$b$^{-1}$ for $p$Ar and $p$Pb respectively, are high
(even when one takes into account a small decay branching ratio of $J/\psi$ into leptons which is not included in the table)
\begin{table}[]
\centering
\begin{tabular}{|c|c|c|}\hline
 Cut on $M_X$ &  $M_X \le 10\, GeV$ &  $M_X \le \sqrt {0.05W_{\gamma p}^2 +t}$\\ \hline
p - Ar  &     0.6 $\mu b$   & 1.2 $\mu b$    \\ \hline
p - Pb  &  9 $\mu b$ & 18 $\mu b$  \\ \hline
\end{tabular}
\caption{Total cross sections of the large rapidity gap $J/\psi$
photoproduction in ultraperipheral p-Ar and p-Pb collisions.}
\label{tcs}
\end{table}

One can see that cross sections are large in the rapidity range where the LHC
detectors have good acceptance to $J/\psi$'s with transverse momenta of few GeV 
and that with the cuts natural for LHC there is a strong sensitivity to the 
value of the $\delta$ parameter.

 Hence we conclude that measurements of the discussed processes in pA 
collisions at LHC would allow to measure energy and momentum transfer 
dependence of the hard vacuum exchange.

\section{Conclusions} \bigskip
 
Studies of rapidity gap processes in the proton-ion UPCs at the LHC will directly measure 
the energy dependence of the large-$t$ elastic amplitude of dipole-parton 
scattering in the wide range of energies. 
We find that it will be possible to study the $J/\psi$ photoproduction in such processes
 up to $W_{\gamma p}\sim 1$ 
TeV, extending the energy range studied at HERA by a factor of 10. This should allow us 
to investigate hard QCD dynamics up to $xs_{\gamma p}/|t| \sim 10^{5}$ that 
corresponds to rapidity interval of $\sim 8$ units for gluon emission.
The rapidity interval between two gluons on the ladder is 
$\ge 2$,
hence, the emission of several gluons is possible in the ladder in such processes
thus making applicability of the BFKL approach more justified. 
These measurements will be also important as a reference
 for the study of a similar process in the nucleus-nucleus
 collisions at the LHC which will be considered in a separate publication
(for discussion of the analogous process of large t  
$\rho$-meson production in the nucleus-nucleus collisions see \cite{yelpage}).

\section*{Acknowledgments}
This work was supported in part by the US Department of
Energy, Contract Number
DE-FG02-93ER40771 (M.~Strikman, M.~Zhalov); L.Frankfurt and MS thank BSF for support;
 M. Zhalov would like to express acknowledgment
for support by CERN-INTAS grant no 05-103-7484 .

\end{document}